\documentclass[aps,pre,twocolumn,groupedaddress,showpacs,amsmath,amssymb]{revtex4-1}

\usepackage{graphicx}

\begin{document}

\def\ldotsplus{\mathinner{\ldotp\ldotp\ldotp\ldotp}}
\def\fourdots{\relax\ifmmode\ldotsplus\else$\m@th \ldotsplus\,$\fi}








\title{Locomotion and transport in a hexatic liquid crystal}










\author{Madison S. Krieger$^1$}
\author{Saverio E. Spagnolie$^2$}
\author{Thomas R. Powers$^{1,3}$}





\affiliation{$^1$School of Engineering, Brown University, Providence, RI 02912 USA}
\affiliation{$^2$Department of Mathematics, University of Wisconsin, Madison WI 53706 USA}
\affiliation{$^3$Department of Physics, Brown University, Providence, RI 02012 USA}







\date{\today}

\begin{abstract}
The swimming behavior of bacteria and other microorganisms is sensitive to the physical properties of the fluid in which they swim. 
Mucus, biofilms, and artificial liquid-crystalline solutions are all examples of fluids with some degree of anisotropy that are also commonly encountered by bacteria. In this article, we study how liquid-crystalline order affects the swimming behavior of a model swimmer. The swimmer is a one-dimensional version of G. I. Taylor's swimming sheet: an infinite line undulating with  small-amplitude transverse or longitudinal traveling waves. The fluid is a two-dimensional hexatic liquid-crystalline film. We calculate the power dissipated, swimming speed, and flux of fluid entrained as a function of the swimmer's waveform as well as properties of the hexatic film, such as the rotational and shear viscosity, the Frank elastic constant, and the anchoring strength. The departure from isotropic behavior is greatest for large rotational viscosity and weak anchoring boundary conditions on the orientational order at the swimmer surface.  We even find that if the rotational viscosity is large enough, the transverse-wave swimmer moves in the opposite direction relative to a swimmer in an isotropic fluid. 
\end{abstract}






\maketitle

\section{Introduction}

Bacteria and other swimming microorganisms often encounter complex fluids which are full of polymers. Mucus is a prime example. Since the polymers are typically rod-shaped and aligned, these fluids can be anisotropic~\cite{TampionGibbons1962,VineyHuberVerdugo1993,Smalyukh_etal2008,FlemmingWingender2010}. 
Furthermore, several experimental groups have recently studied swimmers in synthetic nontoxic liquid-crystalline solutions~\cite{Cheng_etal2005,Shiyanovskii_etal2005,MushenheimEtAl2013,Zhou_etal2013}. These liquid-crystalline solutions are fluids consisting of rod-like molecules that spontaneously align in the absence of external fields~\cite{deGennesProst}.
Liquid crystals are simpler models for the complex anisotropic biological environments encountered by swimming microorganisms, and the anisotropy leads to qualitatively new swimming phenomena not present in isotropic fluids. For example, elastic forces in an liquid crystal can cause bacteria to form multi-cellular assemblies~\cite{MushenheimEtAl2013}. Furthermore, orientation order, both uniform~\cite{MushenheimEtAl2013} and non-uniform~\cite{Zhou_etal2013}, can guide the trajectories of swimming bacteria.   

In this article we explore the effects of elasticity and orientational order on swimming with a simple theory for an idealized microorganism in a two-dimensional hexatic liquid crystal.  
\begin{figure}[b]
\includegraphics[width=3.3in]{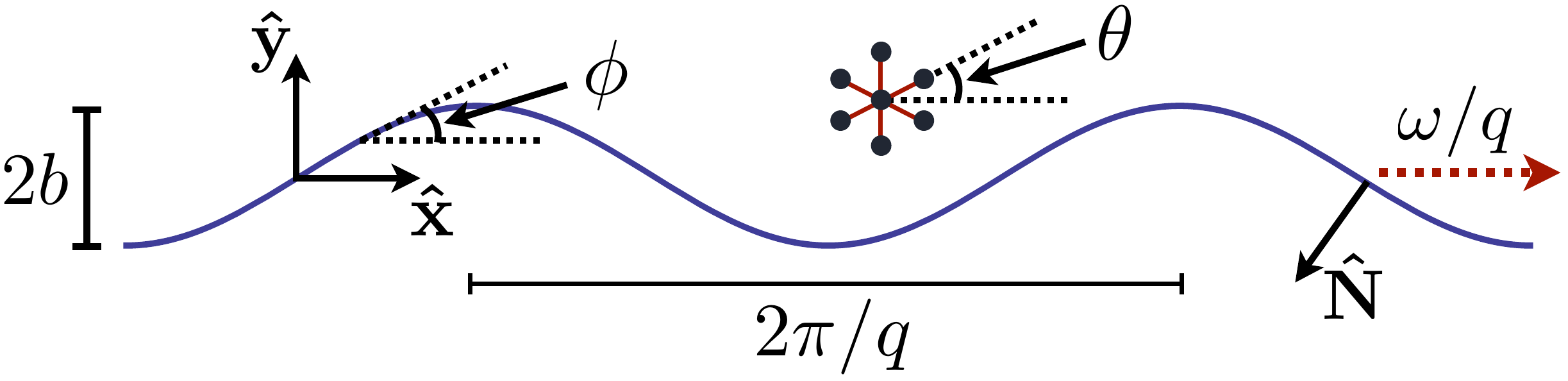}
\caption{(Color online)
Illustration of a swimming sheet immersed in a hexatic liquid crystal (not to scale).
The propagating transverse wave has wavevector $q$, frequency $\omega$, and amplitude $b\ll2\pi/q$. The angle field $\theta$ is defined up to rotations by $2\pi/6$ by the angle between the $x$-axis and any of the lines connecting a particle at $(x,y)$ with its six nearest neighbors.}
\label{setup}
\end{figure}
Our model for the swimmer is a one-dimensional Taylor swimming sheet, an infinite line with internally-generated transverse or longitudinal waves (Fig.~\ref{setup})~\cite{taylor1951}. The fluid is a two-dimensional hexatic liquid crystal film. We work in two dimensions to simplify our calculations.  Furthermore, bacteria and other swimming microorganisms are often studied in quasi-two-dimensional environments, such as thin layers of fluid~\cite{Guasto:2010dz,KurtulduGuastoJohnsonGollub2011,SwiecickiSilusarenkoWeibel2013} or soap films~\cite{wu00}.
We study the  hexatic phase because it is the simplest liquid-crystalline phase~\cite{deGennesProst}, yet it shares many features with the nematic phase encountered by swimming microorganisms~\cite{MushenheimEtAl2013,Zhou_etal2013}, such as orientational elasticity and anchoring effects.  In a hexatic phase, the spontaneous alignment can be visualized by considering the six nearest neighbors of each particle (Fig.~\ref{setup}). On average, these six nearest neighbors define three axes, which  define an imaginary hexagon around each particle. In a hexatic liquid crystal, the centers of mass of the hexagons are disordered, but the orientations of the hexagons share a common alignment. The alignment is described by an angle field $\theta$ (Fig.~\ref{setup}).

The presence of the hexatic order leads to several properties not present in an isotropic fluid. There are elastic torques that tend to drive the system to a state of uniform alignment; these torques are characterized by an elastic constant $K$.
In addition to the usual shear viscosity $\mu$ of the fluid, there is another viscosity $\gamma$ that arises when hexatic order is present. This additional viscosity coefficient
 characterizes the dissipation that arises when the local rate of rotation of the hexagons differs from the local rate of rotation of the fluid. A measure of the relative importance of viscous and elastic effects in a hexatic liquid crystal is the Ericksen number~\cite{larson1999}
\begin{equation}\mathrm{Er}=\frac{\mu\omega}{Kq^2},\label{Erdef}
\end{equation}
where $\omega$ is the beat frequency of the swimmer and $q$ is the wavenumber (Fig.~\ref{setup}).
Although the beat frequencies of undulating cilia and flagella vary widely, typical values are $\omega\approx100\,$s$^{-1}$ and $q\approx1\,\mu$m$^{-1}$.  The disodium cromoglycate (DSCG) liquid crystalline solution used in experiments with swimming bacteria has a viscosity $\mu\approx 1$\,Pa-s~\cite{MushenheimEtAl2013} and elasticity $K\approx 10$\,pN~\cite{MushenheimEtAl2013}. These values lead to $\mathrm{Er}\approx10$. Note the sensitive dependence of Er on the length scale $q^{-1}$:  increasing or decreasing $q$ by a factor of ten can easily put the swimmer in the regimes where elastic or viscous effects dominate, respectively. Using the size and velocity of the swimming bacteria~\cite{MushenheimEtAl2013} to define the Ericksen number leads to Er$<1$. 

The boundary conditions on the angle $\theta$ near the surface of the swimmer are also important, and are governed by an anchoring potential of strength $W$~\cite{RapiniPapoular1969}. For a \textit{two}-dimensional liquid crystal film, the anchoring strength leads to a length scale, $K/W$~\cite{Lavrentovich2014}. For a static undulation of wavenumber $q$, the angle field $\theta$ is uniform throughout the liquid crystal when the anchoring is weak, $w\equiv W/(Kq)\ll1$. The angle field has modulation with wavenumber $q$ when the anchoring is strong, $w\gg1$~\cite{deGennesProst}. 

\section{Hexatic dynamics}
A hexatic liquid crystal has six-fold  bond-orientational order~\cite{deGennesProst}. Symmetry of the director field $\hat{\mathbf{n}}=(\cos\theta,\sin\theta)$ under the rotations $\theta\mapsto\theta+2\pi/6$ (Fig.~\ref{setup}) rules out the splay [$(\nabla\cdot\mathbf{n})^2$] and bend [$(\boldsymbol{\nabla}\times\mathbf{n})^2$] terms of the two-dimensional nematic free energy,  leaving a single bulk term with elastic constant $K$. The full hexatic free energy is
\begin{equation}
F=\frac{K}{2}\int(\nabla\theta\cdot\nabla\theta)\mathrm{d}x\mathrm{d}y+\frac{W_6}{2}\int\sin^2[6(\theta-\phi)]\mathrm{d}\ell,
\label{hexenergy}
\end{equation}
where the first integral is over the domain of the fluid, the second integral is over the boundary of the swimmer, and $W_6$ is the strength of an anchoring potential~\cite{RapiniPapoular1969} that gives a preference for $(\cos\theta,\sin\theta)$ to align with the tangent vector $\hat{\boldsymbol{\ell}}=(\cos\phi,\sin\phi)$ of the  boundary (Fig.~\ref{setup}). Since we consider small-amplitude waves only, we may expand the anchoring term for small angle and absorb the factors of 6 into a new coefficient $W=36W_6$:
\begin{equation}
\frac{W_6}{2}\int\sin^2[6(\theta-\phi)]\mathrm{d}\ell\approx \frac{W}{2}\int^\infty_{-\infty}(\theta-\phi)^2\mathrm{d}x.
\end{equation}
The stress takes the form
\begin{eqnarray}
\sigma_{ik}=&-&p\delta_{ik}+\mu\left(\partial_iv_k+\partial_kv_i\right)\nonumber\\
&-&K\partial_i\theta\partial_k\theta+\frac{K}{2}\epsilon_{ik}\nabla^2\theta,
\end{eqnarray}
where $p$ is the pressure. The elastic part of the stress may be derived from the  free energy using the principle of virtual work, along with the condition that the angle field rotates with the local rotation of the fluid under a virtual displacement~\cite{deGennesProst}.
Since inertia is irrelevant at the scale of microorganisms, we work in the limit of  zero Reynolds number, $\mathrm{Re}=0$, where $\mathrm{Re}=\rho\omega/(\mu q^2)$ is zero and $\rho$ is the density of the fluid~\cite{purcell1977}. Thus, conservation of momentum becomes force balance $\partial_k\sigma_{ik}=0$, which leads to
\begin{equation}
-{\bm \nabla}p+\mu\nabla^2{\mathbf v}-K{\bm\nabla}\cdot\left({\bm \nabla}\theta{\bm\nabla}\theta\right)+\frac{K}{2}{\bm\nabla}\times\left(\hat{\mathbf z}\nabla^2\theta\right)={\mathbf 0},\label{hexfbal}
\end{equation} 
where $\hat{\mathbf z}$ is the unit vector perpendicular to the film, and the pressure $p$ is chosen to enforce ${\bm\nabla}\cdot\mathbf{v}=0$.
Varying the free energy~(\ref{hexenergy}) with respect to $\theta$ yields the equilibrium condition $\nabla^2\theta=0$ which together with Eqn.~(\ref{hexfbal}) implies that the pressure at equilibrium is $p_\mathrm{eq}=-(K/2)\partial_k\theta\partial_k\theta$. 
We can simplify 
Eq.~(\ref{hexfbal}) somewhat by writing $p$ as the sum of the dynamic and equilibrium pressure, $p=p_\mathrm{dyn}+p_\mathrm{eq}$:
\begin{equation}
-{\bm \nabla}p_\mathrm{dyn}+\mu\nabla^2{\mathbf v}-K({\bm\nabla}\theta)\nabla^2\theta+\frac{K}{2}{\bm\nabla}\times\left(\hat{\mathbf z}\nabla^2\theta\right)={\mathbf 0}.
\end{equation}
Henceforth we use $p$ to denote the dynamic pressure. 
Note that the pressure $p$ need not be a harmonic function, as it must for Stokes flow of an incompressible isotropic Newtonian liquid.
The dynamical equation for $\theta$ takes the form~\cite{landau_lifshitz_elas}
\begin{equation}
\partial_t \theta+\mathbf{v}\cdot\bm{\nabla}\theta-\frac{1}{2}\hat{\mathbf z}\cdot{\bm \nabla}\times\mathbf{v}=\frac{K}{\gamma}\nabla^2\theta.\label{tbal}
\end{equation}
$\gamma$ is the rotational viscosity and $K/\gamma$ has units of length squared per unit time and acts as a diffusivity for orientational order. It is convenient to choose units that make the governing equations dimensionless. Measuring time in units of $\omega^{-1}$, length in units of $q^{-1}$, and pressure in units of $\mu\omega$ yields
\begin{eqnarray}
-{\bm \nabla}p+\nabla^2{\mathbf v}&=&-\frac{1}{\mathrm{Er}}\left[\frac{1}{2}{\bm\nabla}\times\left(\hat{\mathbf z}\nabla^2\theta\right)-({\bm\nabla}\theta)\nabla^2\theta\right]\label{dimlesshexeqna}\\
\frac{1}{\mathrm{Er}}\frac{\mu}{\gamma}\nabla^2\theta&=&\partial_t \theta+\mathbf{v}\cdot\bm{\nabla}\theta-\frac{1}{2}\hat{\mathbf z}\cdot{\bm \nabla}\times\mathbf{v}.\label{dimlesshexeqn}
\end{eqnarray}
For boundary conditions, we assume that the fluid does not slip relative to the swimmer on its surface, and that the fluid has uniform velocity $\mathbf{v}=U\hat{\mathbf{x}}$ in the frame of the swimmer in the region far from the boundary, $y\rightarrow\infty$. The boundary condition on $\theta$ at the immersed body is deduced by varying the free energy~(\ref{hexenergy}) with respect to $\theta$, which yields 
\begin{equation}
K\hat{\mathbf{N}}\cdot{\bm \nabla}\theta+W(\theta-\phi)=0,\label{anchBC}
\end{equation}
where $\hat{\mathbf N}$ is the outward-pointing normal (Fig.~\ref{setup}), and again $\phi$ is the angle between the tangent vector of the swimmer and the $x$-axis.

\begin{figure}[t]
\begin{center}
\includegraphics[height=2in]{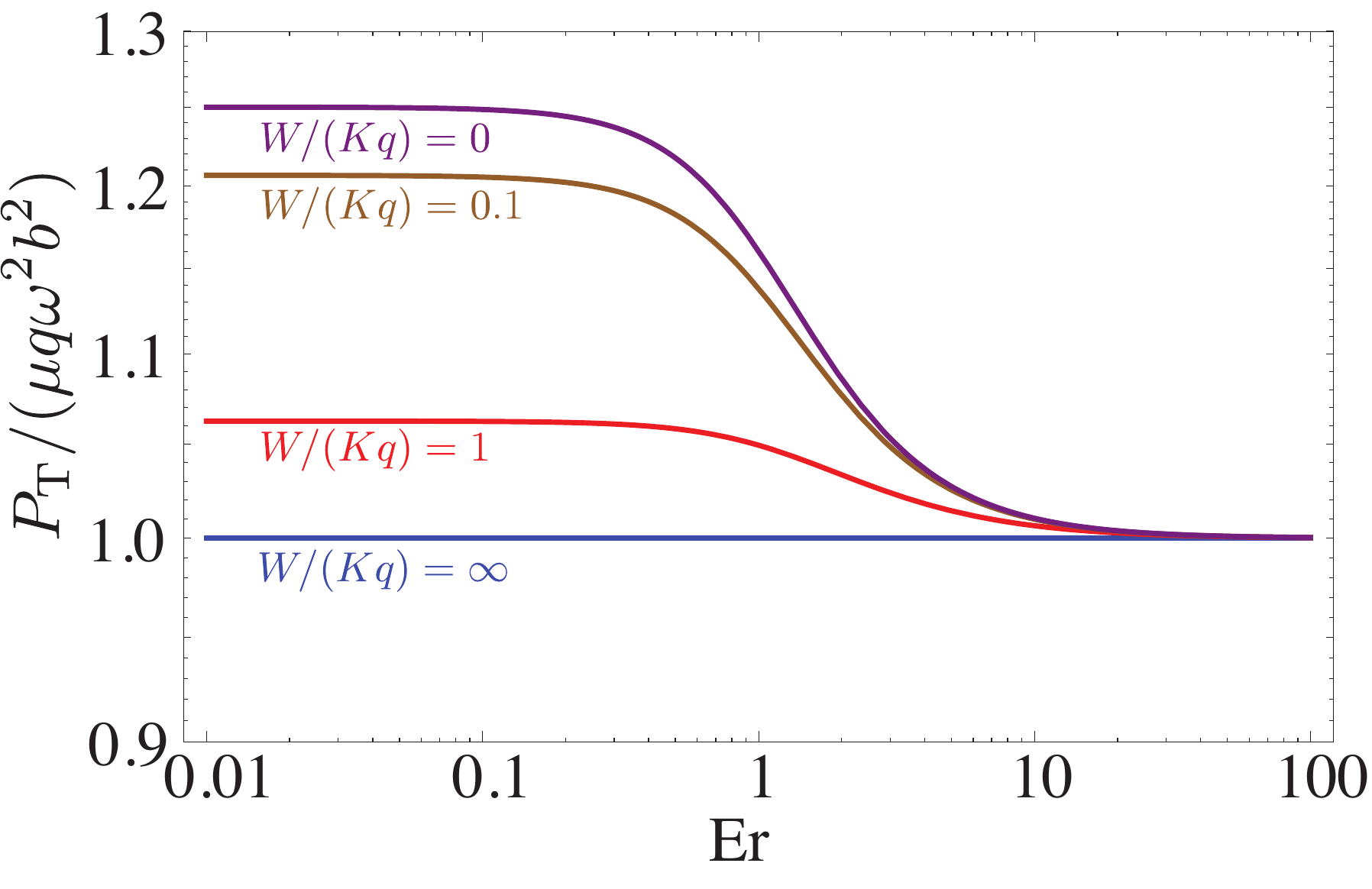}
\caption{(Color online)
Semi-log plot of dimensionless power $P_\mathrm{T}/(\mu q\omega^2 b^2)$ vs. Ericksen number Er for $\gamma=\mu$ and various dimensionless anchoring strengths $w=W/(Kq)$ for a transverse wave.}
\label{Pv}
\end{center}
\end{figure}

\section{Small-amplitude expansion}
 For the swimmer, we consider two kinds of waves: transverse traveling waves 
in which the material points of the swimmer are $(x_\mathrm{s},y_\mathrm{s})=(x,y_1(x,t))$, with $y_1=b\sin(qx-\omega t)$; and longitudinal traveling waves, in which the material points of the swimmer are $(x_\mathrm{s},y_\mathrm{s})=(x+u_1(x,t),0)$, with $u_1=a\sin(qx-\omega t)$. We only consider the cases of a pure transverse or pure longitudinal wave. In both cases the wave propagates rightward, and a positive swimming speed $U>0$ indicates swimming opposite to the direction of wave propagation.  
Following Taylor~\cite{taylor1951}, we expand the fields in powers of the dimensionless amplitude, using a superscript to denote the power of $\varepsilon_b=bq$ ($\varepsilon_a=aq$) for transverse (longitudinal) waves. For example, $\mathbf{v}=\mathbf{v}^{(1)}+\mathbf{v}^{(2)}+\fourdots$

\subsection{First-order equations} It is convenient to express the governing equations in terms of a stream function $\psi$, which is related to the velocity field by $\mathbf{v}=(v_x,v_y)=\bm{\nabla}\times(\psi\hat{\mathbf{z}})$. To first order in amplitude,
\begin{eqnarray}
\nabla^4\psi^{(1)}+\frac{1}{\mathrm{2Er}}\nabla^4\theta^{(1)}&=&0
\label{stokes1stream}\\
\partial_t\theta^{(1)}+\frac{1}{2}\nabla^2\psi^{(1)}&=&\frac{1}{\mathrm{Er}}\frac{\mu}{\gamma}\nabla^2\theta^{(1)}.\label{theta1stream}
\end{eqnarray}
The solutions are given by the real parts of 
\begin{eqnarray}
\hat{\psi}^{(1)}&=&(c_0+c_1 y)\mathrm{e}^{-y+\mathrm{i}(x-t)}-\frac{1}{2\mathrm{Er}} \hat\theta^{(1)}\label{psieqn}\\
\hat{\theta}^{(1)}&=&\left[\mathrm{i}c_1\mathrm{e}^{-y}+c_2 \mathrm{e}^{k y}\right]\mathrm{e}^{\mathrm{i}(x-t)},\label{thetaeqn}
\label{foeqns}
\end{eqnarray}
where 
\begin{equation}
k=-\sqrt{1-\frac{4\mathrm{i}\gamma\mathrm{Er}}{\gamma+4\mu}},\label{keqn}
\end{equation} 
and the constants $c_0$, $c_1$, and $c_2$ are determined by 
the no-slip and anchoring boundary conditions at the surface of the swimmer. For the purely transverse wave, the  no-slip boundary condition $\mathbf{v}(x_\mathrm{s},y_\mathrm{s})=(x_\mathrm{s},\partial_t y_\mathrm{s})$ is
\begin{figure}[t]
\begin{center}
\includegraphics[height=2in]{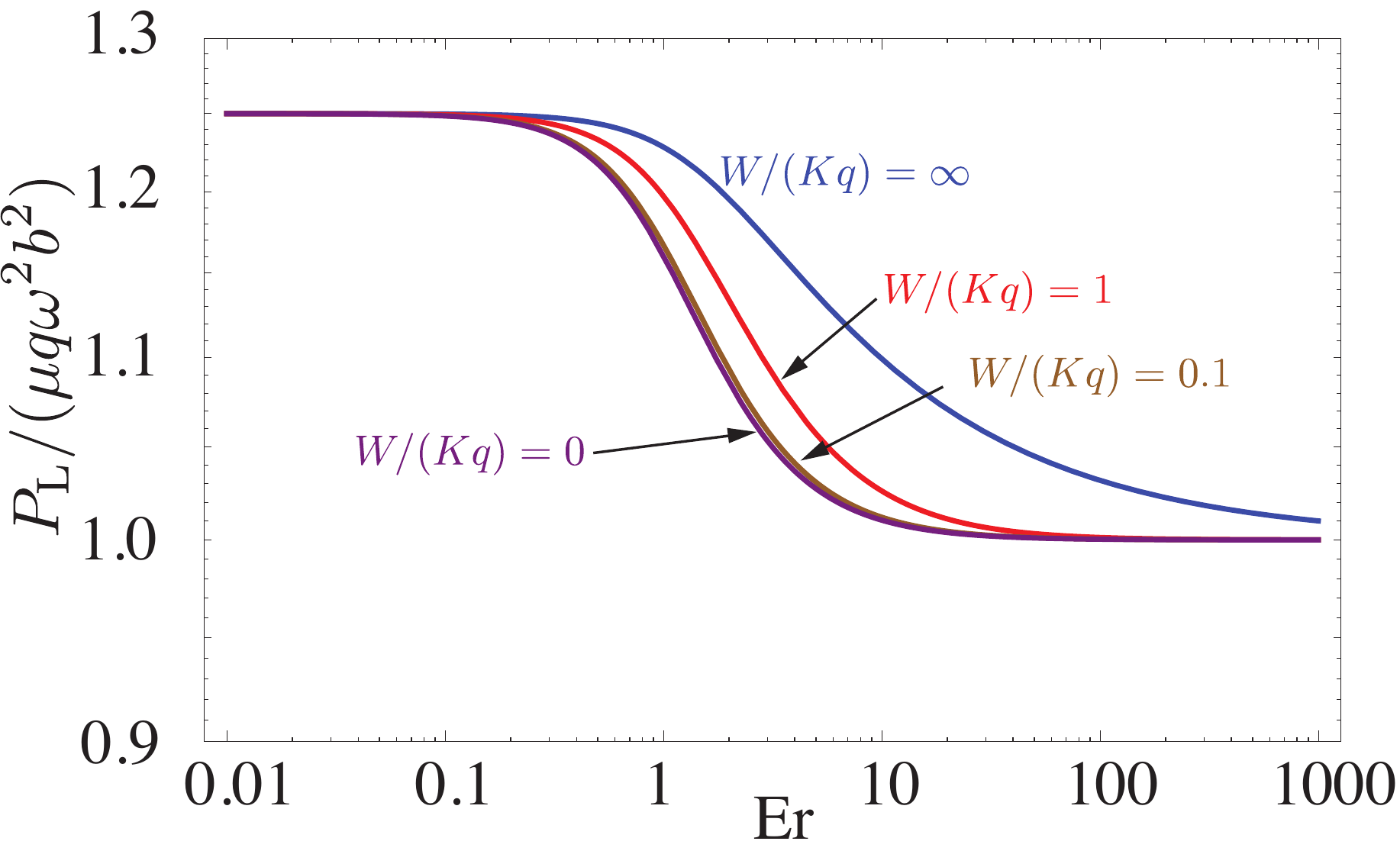}
\caption{(Color online)
Semi-log plot of dimensionless power $P_\mathrm{L}/(\mu q\omega^2 b^2)$ vs. Ericksen number Er for $\gamma=\mu$ and various dimensionless anchoring strengths $w=W/(Kq)$ for a longitudinal wave.}
\label{PvL}
\end{center}
\end{figure}

\begin{equation}
\left.\left(\partial_y\psi^{(1)},-\partial_x\psi^{(1)}\right)\right|_{y=0}=\left(0,-\varepsilon_b\cos(x-t)\right)\label{bc1}
\end{equation}
to first order in dimensionless form. 
Likewise, to first order, the anchoring boundary condition (\ref{anchBC}) is
\begin{equation}
-\left.\partial_y\theta^{(1)}\right|_{y=0}+w\left(\left.\theta^{(1)}\right|_{y=0}-\varepsilon_b\cos(x-t)\right)=0.\label{bc1anch}
\end{equation}
For the purely longitudinal wave, the first-order conditions at the swimmer are 
\begin{eqnarray}
\left.\left(\partial_y\psi^{(1)},-\partial_x\psi^{(1)}\right)\right|_{y=0}&=&(-\varepsilon_a\cos(x-t),0)\label{bc1long}\\
(-\partial_y\theta^{(1)}+w\theta^{(1)})|_{y=0}&=&0\label{bc2long}.
\end{eqnarray}
Analytic expressions for the constants $c_0$, $c_1$, and $c_2$ and the first order quantities $\psi^{(1)}$ and $\theta^{(1)}$ may be found for both the transverse and longitudinal wave; however, these expressions are too unwieldy to display here. We consider the limiting values of large and small Er below. It should be noted that since the solutions~(\ref{psieqn}, \ref{thetaeqn}) have a zero-average in $\hat{\mathbf{x}}$ there is no swimming speed to first order.

We can calculate the power dissipated to second order in amplitude using only the first-order solutions. 
In addition to the shear component familiar from isotropic fluids,  the total power dissipated has a component arising from rotation of the directors relative to the local rotation of the fluid:
\begin{equation}
\mathcal{P}=\int\left(2\mu v_{ij}v_{ij}+h^2/\gamma\right)\mathrm{d}x\mathrm{d}y,\\
\end{equation}
where (in dimensional form) $v_{ij}=(\partial_i v_j+\partial_j v_i)/2$ is the symmetric rate-of-strain tensor and $h=\nabla^2\theta/\mathrm{Er}$ is the molecular field~\cite{landau_lifshitz_elas}. In dimensionless form, the power dissipated is
\begin{equation}
\frac{\mathcal{P}}{\mu\omega^2}=\int\left[2\mu v_{ij}v_{ij}+\frac{1}{\mathrm{Er}^2}\frac{\mu}{\gamma}(\nabla^2\theta)^2\right]\mathrm{d}x\mathrm{d}y.
\end{equation}
The total power dissipated is infinite, since the swimmer is infinite, but we can calculate the power $P$ dissipated per wavelength.  Figures~\ref{Pv} and~\ref{PvL} show the power dissipated for $\mu=\gamma$ and various anchoring strengths for transverse and longitudinal waves, respectively.

Note that when $\gamma=0$, Eqn.~(\ref{tbal}) shows that there is no coupling between the angle field $\theta$ and the flow field $\mathbf{v}$. The angle field takes  the equilibrium configuration given by $\nabla^2\theta^{(1)}=0$, and the first-order flow field is the same as that found by Taylor in the isotropic case
~\cite{taylor1951}. Thus, independent of the form of the wave, the power goes to the isotropic result: $P\approx\mu q\omega^2b^2$ when $\gamma\ll\mu$. Note also that the power approaches the isotropic value when Er is large. This result may be expected since in the limit of $\mathrm{Er}\rightarrow\infty$, the governing equations~(\ref{dimlesshexeqna}--\ref{dimlesshexeqn}) reduce to Stokes equations for an isotropic fluid, with the angle field rotating with the local rate of rotation of the fluid and thus incurring no rotational dissipation. However, we will see below that the $\mathrm{Er}\rightarrow\infty$ limit is singular, and will explore in more detail which quantities approach the isotropic values for large $\mathrm{Er}$.

\subsection{Second-order equations} Now consider the second order equations for $v_x^{(2)}$, averaged over a period. Total derivatives in $x$  or $t$ vanish upon averaging, e.g. $\langle \partial_x p\rangle=0$, leading to
\begin{eqnarray}
\langle \partial^2_yv_x^{(2)}\rangle
+\frac{1}{2\mathrm{Er}}\langle \partial^3_y\theta^{(2)}\rangle&=&f\label{hextf2a}\\
\frac{1}{\mathrm{Er}}\frac{\mu}{\gamma}\langle\partial_y^2\theta^{(2)}\rangle-\frac{1}{2}\langle\partial_y v_x^{(2)}\rangle&=&g,
\label{hextf2b}
\end{eqnarray}
where $f=\langle\partial_x\theta^{(1)}\nabla^2\theta^{(1)}\rangle/\mathrm{Er}$ and $g=\langle\mathbf{v}^{(1)}\cdot\bm{\nabla}\theta^{(1)}\rangle$.
The no-slip boundary condition for a transverse wave is
\begin{equation}
\left.\langle v_x^{(2)}\rangle\right|_{y=0}=-\left.\langle y_1\partial_y v_x^{(1)}\rangle\right|_{y=0},\label{fullvbc}
\end{equation}
whereas for a longitudinal wave we have
\begin{equation}
\left.\langle v_x^{(2)}\rangle\right|_{y=0}=-\left.\langle u_1\partial_x v_x^{(1)}\rangle\right|_{y=0}.\label{nosliplong}
\end{equation}
Note that the right-hand sides of both Eqn.~(\ref{fullvbc}) and (\ref{nosliplong}) depend on anchoring strength through the first-order velocity.
The flow field and thus the swimming speed $U_\mathrm{T}$ is found by solving Eqs.~(\ref{hextf2a}--\ref{hextf2b}) for $\langle v_x^{(2)}\rangle$ subject to the no-slip boundary condition (\ref{fullvbc}) or (\ref{nosliplong}), as well as demanding that $\langle v_x^{(2)}\rangle$ and $\langle\theta^{(2)}\rangle$ be finite at $y\rightarrow\infty$. Independent of whether the wave is transverse or longitudinal, the result is
\begin{equation}
\langle v_x^{(2)}\rangle=\left.\langle v_x^{(2)}\rangle\right|_{y=0}-\frac{4\mu}{4\mu+\gamma}\int_0^y\left(F+\frac{\gamma}{2\mu} g\right)\mathrm{d}y',\label{Utint}
\end{equation}
where $F(y)=\int_y^\infty f(y')\mathrm{d}y'$. Note that the boundary conditions on $\langle\theta^{(2)}\rangle$ do not enter the expression for $\langle v_x^{(2)}\rangle$. The swimming speed $U$ is given by the flow speed at $y=\infty$:
\begin{equation}
U=\left.\langle v_x^{(2)}\rangle\right|_{y=0}-\frac{4\mu}{4\mu+\gamma}\int_0^\infty\left(F+\frac{\gamma}{2\mu} g\right)\mathrm{d}y.\label{Utint2}
\end{equation}

Similarly, the general expression for $\langle\theta^{(2)}\rangle$ is
\begin{equation}
\langle\theta^{(2)}\rangle=\Theta+\frac{\gamma\mathrm{Er}}{2\mu}\int_0^y\left[\langle v_x^{(2)}(y')\rangle-U-2G(y')\right]\mathrm{d}y',
\end{equation}
where $G(y)=\int_y^\infty g(y')\mathrm{d}y'$, and the constant $\Theta$ is determined by the anchoring boundary condition~(\ref{anchBC}). 
For the transverse wave, the second-order part of the anchoring condition (\ref{anchBC}) takes the form
\begin{equation}
\left[-\langle\partial_y\theta^{(2)}\rangle+w\langle\theta^{(2)}\rangle\right]_{y=0}=\Upsilon\label{anchBC2order},
\end{equation}
where $\Upsilon=\Upsilon_\mathrm{T}$ for the transverse wave, and $\Upsilon=\Upsilon_\mathrm{L}$ for the longitudinal wave, with
\begin{eqnarray}
\Upsilon_\mathrm{T}&=&\left.\langle -\partial_xy_1\partial_x\theta^{(1)}+y_1\partial_y^2\theta^{(1)}-wy_1\partial_y\theta^{(1)}\rangle\right|_{y=0}\nonumber\\
\Upsilon_\mathrm{L}&=&\left.\langle  u_1\partial_x\partial_y\theta^{(1)}-w u_1\partial_x\theta^{(1)}\rangle\right|_{y=0}.
\end{eqnarray}
Thus, 
\begin{equation}
\Theta=\frac{\Upsilon}{w}-\frac{2}{w}\frac{\gamma\mathrm{Er}}{4\mu+\gamma}\int_0^\infty(F-2g)\mathrm{d}y'.
\end{equation}


We will be interested in a third observable in addition to the swimming speed of the filament and the angle field. It turns out that unlike the case of the Taylor swimmer in  a Newtonian or Oldroyd-B fluid at zero Reynolds number, there is fluid pumped by a swimmer in a hexatic liquid crystal. In the lab frame, the average flux is given by 
\begin{eqnarray}
Q&=&\int_{y_\mathrm{s}}^\infty\langle v_x-U\rangle\mathrm{d}y\nonumber\\
&\approx&\int_0^\infty\langle v_x^{(2)}-U\rangle
\mathrm{d}y-\left.\langle y_\mathrm{s}v_x^{(1)}\rangle\right|_{y=0}.
\label{flux}
\end{eqnarray}
Note that the second term of Eqn.~(\ref{flux}) vanishes for a transverse wave since $v_x^{(1)}|_{y=0}=0$, and for a longitudinal wave since $y_\mathrm{s}=0$. Therefore, the flux is given to second-order accuracy by 
\begin{equation}
Q^{(2)}=\int_0^\infty(\langle v_x^{(2)}\rangle-U)\mathrm{d}y.
\end{equation}
Note that our sign convention for flux is opposite that for swimming speed: a positive $U$ means swimming to the left, whereas a positive $Q$ means fluid is swept to the right. 
\begin{figure*}[t]
\begin{center}
\includegraphics[height=1.65in]{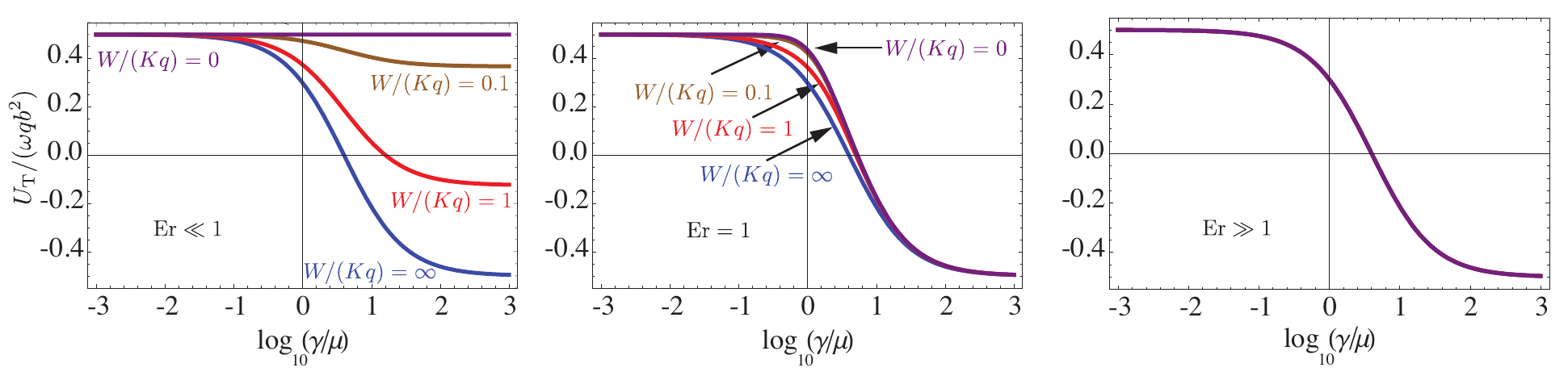}
\caption{(Color online)
Semi-log plot of dimensionless swimming speed $U/(\omega q b^2)$ vs. $\gamma/\mu$ for Er$\ll1$ (left panel), Er$=1$ (middle panel), and Er$\gg1$ (right panel), for a transverse wave. Note that the strong anchoring case $W/(Kq)=\infty$ is independent of Er, and that anchoring effects are small once Er is of order unity. }
\label{Utvg}
\end{center}
\end{figure*}

\begin{figure}[b]
\begin{center}
\includegraphics[height=2.0in]{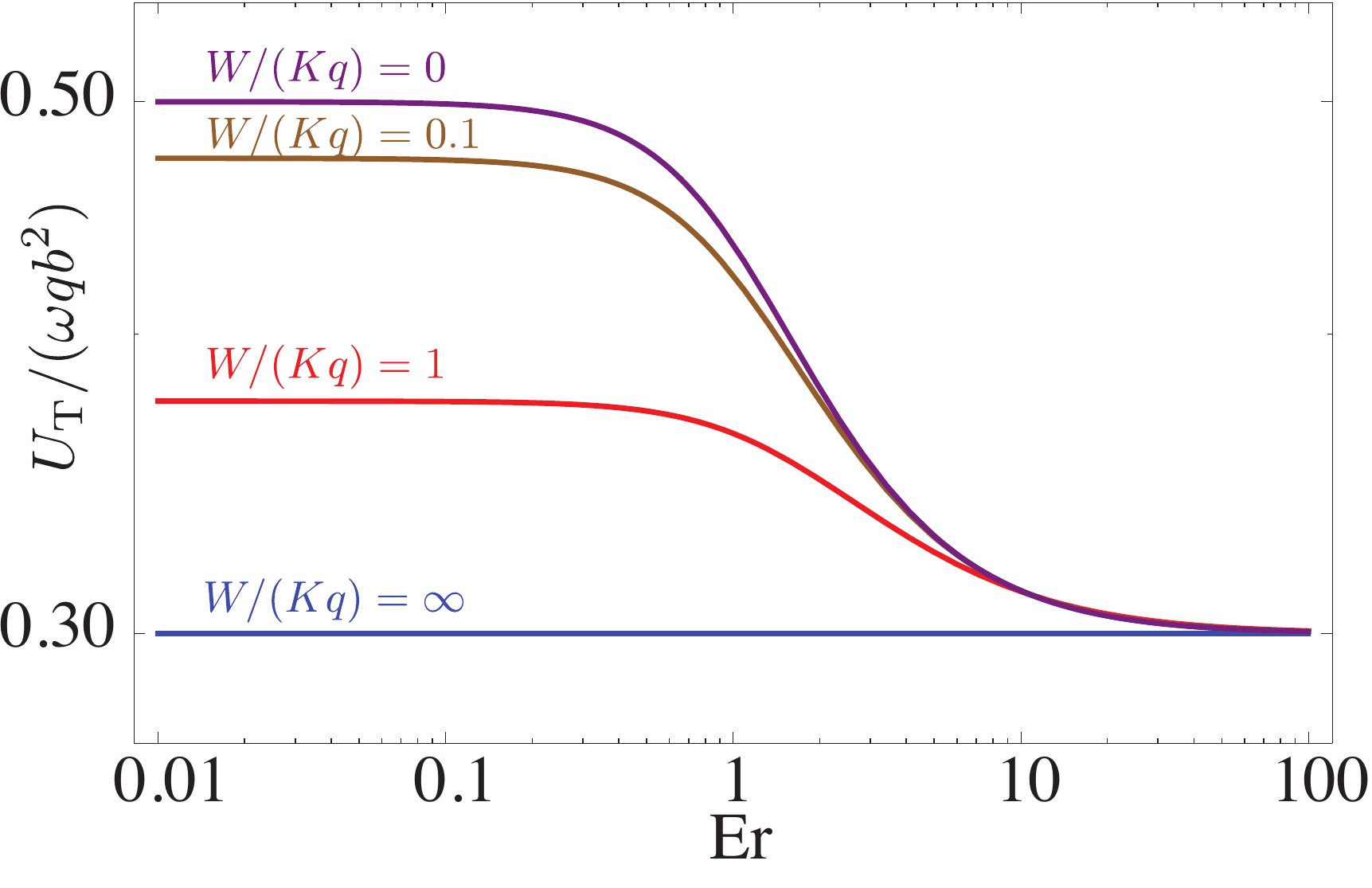}
\caption{(Color online)
Log-log plot of dimensionless swimming speed $U/(\omega q b^2)$ for a transverse wave with $\gamma=\mu$ and various dimensionless anchoring strengths $W/(Kq)$.}
\label{UvEr}
\end{center}
\end{figure}

\begin{figure}[b]
\begin{center}
\includegraphics[height=2.0in]{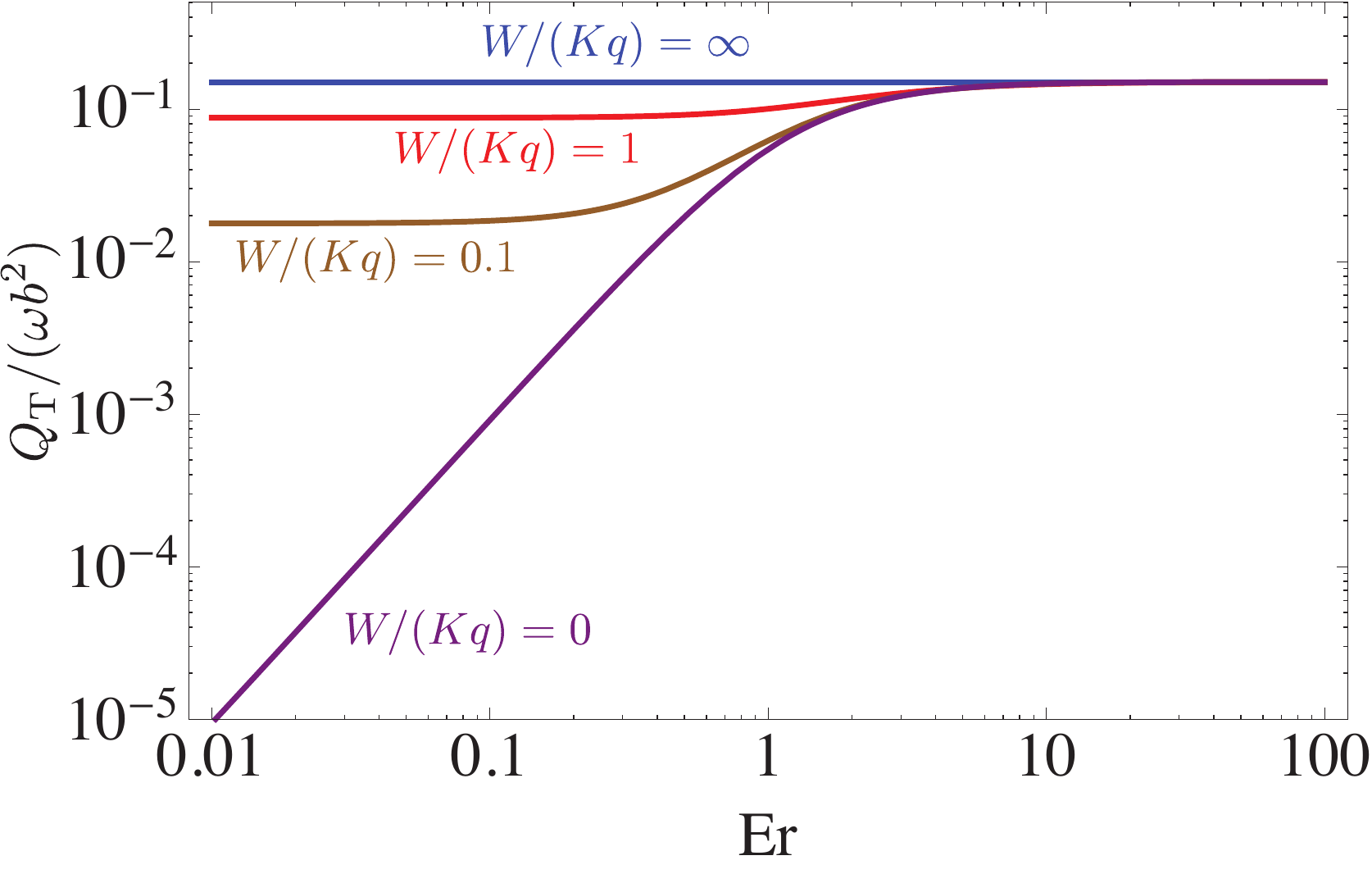}
\caption{(Color online)
Log-log plot of dimensionless flux $Q/(\omega b^2)$ for a transverse wave with $\gamma=\mu$ and various dimensionless anchoring strengths $W/(Kq)$.}
\label{QtvErg1}
\end{center}
\end{figure}

\subsection{Results for general values of the parameters} As mentioned above, the analytic expressions for the power, swimming speed, and fluid transported for general values of the parameters are too complicated to display. However, it is straightforward to plot these quantities as function of the ratio of rotational and shear viscosity $\gamma/\mu$ as well as Ericksen number. 
Figure~\ref{Utvg} shows how the speed $U_\mathrm{T}$ of a swimmer with a transverse wave depends on $\gamma/\mu$ for various anchoring strengths at small, intermediate, and large Ericksen numbers. 
When $\gamma\ll\mu$, the swimming speed is the same as the isotropic speed $\omega q b^2/2$ for all $\mathrm{Er}$ and all anchoring strengths, since there is no coupling between the hexatic degrees of freedom and flow when $\gamma=0$. The speed depends on the anchoring strength only when $\gamma>\mu$ and the Ericksen number is small. When $\mathrm{Er}\gtrsim1$, the speed is only weakly dependent on $w$, becoming independent of $w$ when $\mathrm{Er}\gg1$. The swimming speed at $\mathrm{Er}\gg1$ is different from the isotropic value $\omega q b^2/2$~\cite{taylor1951}. Thus, the large Er limit is singular, since when $\mathrm{Er}=\infty$, the governing equation for flow (\ref{dimlesshexeqna}) reduces to Stokes equation, which leads $U_\mathrm{T}=\omega q b^2/2$.

Figures~\ref{UvEr} and~\ref{QtvErg1} show how the speed and flux depend on Ericksen number, respectively, for a swimmer with a transverse wave and $\gamma=\mu$.  Just as for the power dissipated, the speed and flux depend on anchoring conditions only for small Ericksen number, where elastic stresses are larger than viscous stresses. When the anchoring strength is weak, the swimming speed and the flux go to the isotropic results at small Er, with the flux $Q_\mathrm{T}\propto\mathrm{Er}^2$ when $w=0$. At large Ericksen number, the speed and flux are independent of the anchoring strength, and different from the isotropic values, again consistent with the singular limit $\mathrm{Er}\rightarrow\infty$. 

\begin{figure*}[t]
\begin{center}
\includegraphics[width=6.8in]{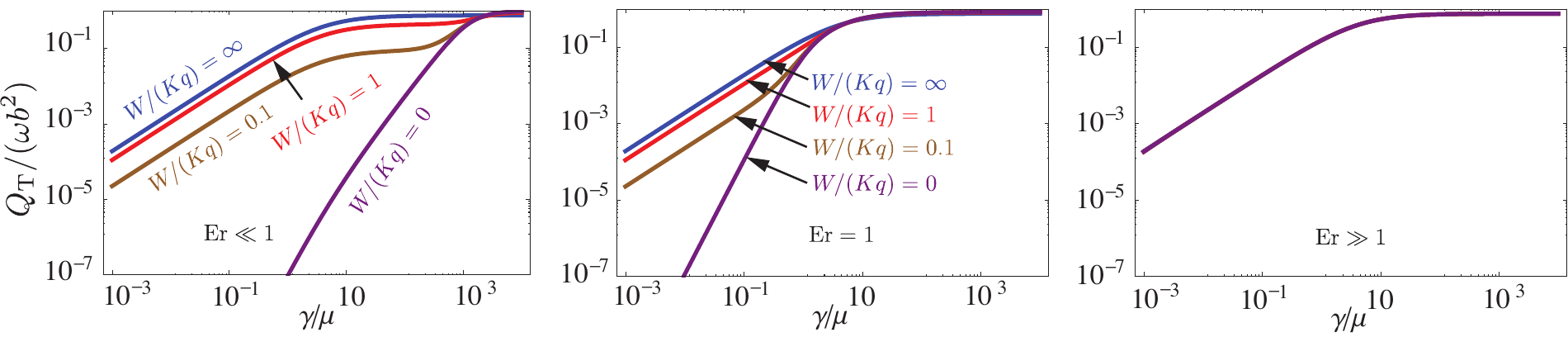}
\caption{(Color online)
Log-log plot of dimensionless flux $Q/(\omega  b^2)$ vs. $\gamma/\mu$ for a transverse wave for Er$\ll1$ (left panel), Er$=1$ (middle panel), and Er$\gg1$ (right panel). Note that the strong anchoring case $W/(Kq)=\infty$ is independent of Er, and the flux is very small for all cases when $\gamma/\mu<0.1$. }
\label{Qtvg}
\end{center}
\end{figure*}

\begin{figure}[b]
\begin{center}
\includegraphics[height=2.0in]{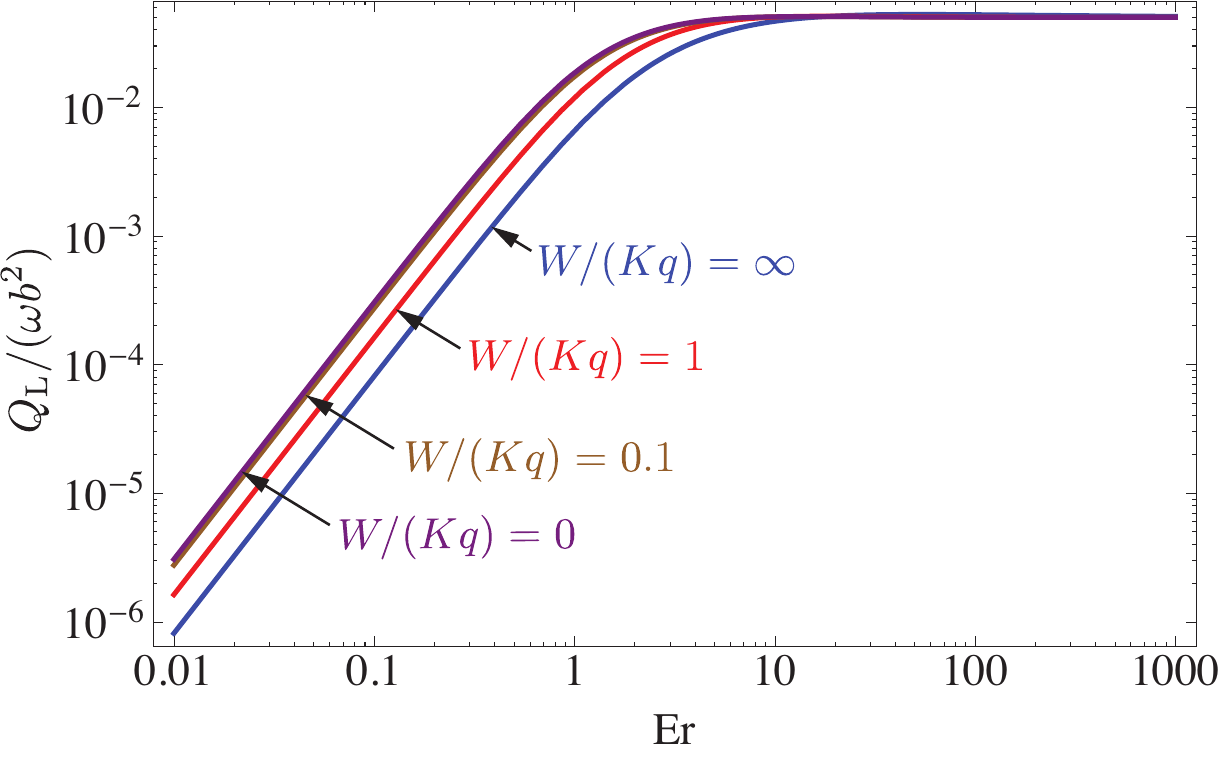}
\caption{(Color online)
Dimensionless flux for a longitudinal swimmer vs. Ericksen number Er for $\gamma=\mu$ and various dimensionless anchoring strengths $W/(Kq)$.}
\label{QLvEr}
\end{center}
\end{figure}

The flux vs. $\gamma/\mu$ for a transverse wave is shown for small and intermediate Er for various anchoring strengths in the left and middle panels of Fig.~\ref{Qtvg}, respectively, and again the dependence on anchoring strength is evident, with $Q_\mathrm{T}\propto\gamma$ for small rotational viscosity, $\gamma\ll\mu$. For large rotational viscosity, the flux approaches an Ericksen-number-indepenent value. Once again, at large Er (Fig.~\ref{Qtvg}, right panel), we see that anchoring strength does not affect the flux.

We do not plot the speed of a swimmer with a longitudinal wave, since it is always within a few percent of the isotropic longitudinal speed $U_\mathrm{L}=-\omega q b^2/2$. However, there is a non-vanishing flux for the longitudinal case large Ericksen number (Fig.~\ref{QLvEr}).  The dependence of $U_\mathrm{L}$ on the anchoring strength is also very weak, just as for the power. For small Er, the flux generated by a longitudinal wave vanishes like $\mathrm{Er}^2$ (Fig.~\ref{QLvEr}). There is a weak dependence on the anchoring strength. At large Er, the flux is independent of $w$ and Er. 




With the swimming speeds and power dissipated in hand, we can calculate measures of efficiency.
However, we cannot define the efficiency as $e=U^2/P$, as is commonly done for swimmers in an isotropic fluid~\cite{LaugaPowers2009}.  This definition rests on the assumption that that the power required to drag an object through a fluid is proportional to $U^2$, with the proportionality constant equal to viscosity times a function of geometrical factors. In the presence of hexatic order, the proportionality constant also depends on the Ericksen number. Thus, $U^2/P$ does not accurately reflect the ratio of the power required to drag the swimmer to the power expended by the swimmer. However, the swimming economy $U/P$ vs. Er is a meaningful quantity. The dimensionless power is close to unity over the entire range of Er for both transverse and longitudinal waves. Thus, the swimming economy has roughly the same form as $U$ vs. Er, decreasing monotonically with Er and approaching a $w$-independent limit at high Er.

To elucidate these results, we turn to a discussion of the asymptotic regimes of the parameter values, for which the calculations and expressions simplify greatly.

\section{Discussion of asymptotic results} We consider the limits of strong anchoring for the transverse wave as well as small and large Ericksen number for both the transverse wave and the longitudinal wave. 
\subsection{Strong anchoring---transverse wave}
The boundary conditions~(\ref{bc1}--\ref{bc1anch}) that determine the coefficients in~(\ref{foeqns}) greatly simplify in the case of a transverse wave with strong anchoring, $W/(Kq)\rightarrow\infty$, leading to 
(dimensionless) solutions \begin{eqnarray}
\theta^{(1)}&=&\varepsilon_b\exp(-y)\cos(x- t)\label{theta1sol}\\
v^{(1)}_x&=&-\varepsilon_by\exp(-y)\sin(x- t)\label{vx1sol}\\
v^{(1)}_y&=&-\varepsilon_b(1+y)\exp(-y)\cos(x-t)\label{vy1sol}\\
p^{(1)}&=&-2\varepsilon_b\exp(-y)\cos(x-t).\label{p1sol}
\end{eqnarray}
Note that since the angle field is harmonic to first order, there are no elastic torques or forces acting on the hexatic and the angle field rotates with the local rate of rotation of the fluid. The flow is the same as  Stokes flow of an isotropic liquid, as found by Taylor~\cite{taylor1951}.  There is no swimming speed to first order in $bq$. 
It is curious that the first-order flow field and angle field are both independent of Ericksen number (as $c_2=0$). 
In contrast, the angle and flow fields depend on Er for the longitudinal swimmer for all anchoring strengths, and for the transverse swimmer for finite anchoring strength.

Turning now to the second-order equations~(\ref{hextf2a}--\ref{hextf2b}),
we saw in Eqn.~(\ref{theta1sol}) that $\theta^{(1)}$ is harmonic, which implies that the right-hand side of Eq.~(\ref{hextf2a}) vanishes; likewise $\langle\mathbf{v}^{(1)}\cdot\bm{\nabla}\theta^{(1)}\rangle=\varepsilon_b^2(1+2y)\exp(-2y)$. We solve Eqs.~(\ref{hextf2a}--\ref{hextf2b}) 
subject to the no-slip and strong anchoring boundary conditions at the swimmer. Expanding to second order in $q^2b^2$, these boundary conditions are
$\langle v_x^{(2)}\rangle(y=0)=\varepsilon_b^2/2$ and
$\langle\theta^{(2)}\rangle(y=0)=0$.
At $y\rightarrow\infty$ we demand that $\langle\theta^{(2)}\rangle$ and $\langle v^{(2)}\rangle$ be finite.
The (dimensionless) solutions to the coupled equations (\ref{hextf2a}--\ref{hextf2b}) are
\begin{eqnarray}
\langle \theta^{(2)}\rangle&=&\frac{\gamma \varepsilon_b^2 \mathrm{Er}}{2(\gamma+4\mu)}\left[(3+2y)\mathrm{e}^{-2y}-3\right]\label{theta2trans}\\
\langle v_x^{(2)}\rangle&=&\frac{\gamma \varepsilon_b^2(1+y)\mathrm{e}^{-2y}}{\gamma+4\mu}+\frac{\varepsilon_b^2(4\mu-\gamma)}{2(4\mu+\gamma)}\label{v2hextrans}.
\end{eqnarray}
The swimming speed is obtained from the limit at $y\rightarrow\infty$. Therefore, in the lab frame, the swimmer with a transverse wave 
swims to the left with (dimensional) velocity
\begin{equation}
U_\mathrm{Ta}=\frac{\omega q b^2}{2}\frac{4\mu-\gamma}{4\mu+\gamma},\label{Uta}
\end{equation}
where $b^2\omega q/2$ is Taylor's result for swimming in a Newtonian Stokes flow~\cite{taylor1951}, the subscript ``T" denotes transverse, and the subscript ``a" denotes strong anchoring. The swimming speed for a transverse wave does not depend on Er, but the direction of swimming depends on $\gamma/\mu$. There is even a point ($4\mu=\gamma$) where the swimmer makes no progress. Just as we observed when discussing the power dissipated, when $\gamma/\mu\rightarrow0$, $U_\mathrm{Ta}$ approaches the result for a swimmer in an isotropic fluid.  
When $\gamma/\mu\rightarrow\infty$, the speed is the same as in the isotropic case, but the direction has reversed. The dependence of the swimming speed on $\gamma/\mu$ for $W/(Kq)=\infty$ is shown in Fig.~\ref{Utvg} (blue curves).

The flow induced by the swimmer has a flux 
\begin{equation}
Q_\mathrm{Ta}=\frac{3}{4}\frac{\gamma\omega b^2}{\gamma+4\mu}.
\end{equation}
The traveling waves on the swimmer move to the right in the swimmer frame, by definition, and also in the lab frame, since $U\ll \omega/q$.  For transverse waves and strong anchoring, the fluid is always pumped in the direction of motion of traveling waves, independent of the direction of swimming. Since the swimming speed is ${\mathcal O}(\omega b^2 q)$, the form of $Q$ implies that the thickness of the layer of fluid swept along by the swimmer is one wavelength $\sim1/q$.

Our solution Eq.~(\ref{theta2trans}) for the angle field $\theta$ has some unexpected features. First, note that when $y\rightarrow\infty$, the swimming-induced disturbance to the director field does \textit{not} vanish. There is a  nonzero $\mathcal{O}(b^2q^2)$ value for the angle field. The solution does not allow us to demand $\langle\theta^{(2)}(y\rightarrow\infty)\rangle\rightarrow0$; we can only demand that the director field be finite at $y\rightarrow\infty$. Just as the flow velocity has a   constant term at $y\rightarrow\infty$, the angle field has a constant term at $y\rightarrow\infty$. The other unexpected feature of the solution is that $\langle\theta^{(2)}\rangle$ is proportional to $\mathrm{Er}$. Thus, our small-amplitude expansion is valid for fixed $\mathrm{Er}$; it is not uniformly valid for large $\mathrm{Er}$. Inspection of Eqns.~(\ref{dimlesshexeqna}) and (\ref{dimlesshexeqn}) might suggest that the flow is isotropic Stokes flow with the angle field rotating at the half the rate of the local vorticity. However, the divergence of $\theta$ with $\mathrm{Er}$ makes the terms involving $\mathrm{Er}\nabla^2\theta$ singular perturbations. Therefore, the limit of the swimming speed at large Er is different from Taylor's result for an isotropic fluid at infinite $\mathrm{Er}$. This situation is similar to the case of a swimmer in an isotropic Newtonian fluid with inertia, in which the inviscid limit of the swimming flow is different from the inviscid flow~\cite{Childress2008}. 

\subsection{Small Ericksen number}
\subsubsection{Transverse wave} In this limit the viscous stresses are weak compared to elastic stresses. At each order in amplitude, we expand in Ericksen number, denoting the power of Er by a subscript. For example, $\psi^{(1)}=\psi^{(1)}_0+\mathrm{Er}\psi^{(1)}_1+\fourdots$ To zeroth order in Er, the equations (\ref{stokes1stream}--\ref{theta1stream}) become 
\begin{eqnarray}
\nabla^2\theta^{(1)}_0&=&0\label{theta10a}\\
\nabla^4\theta^{(1)}_0&=&0.\label{theta10b}
\end{eqnarray}
With the anchoring boundary condition~(\ref{bc1anch}), these imply
\begin{equation}
\theta^{(1)}_0=\varepsilon_b\frac{w}{1+w}\mathrm{e}^{-y}\cos(x-t).
\end{equation}
Note that the large $w$ limit of $\theta^{(1)}_0$ is equal to the angle field we found in the strong anchoring case, Eqn.~(\ref{theta1sol}).
To first order in Er, Eqs.~(\ref{stokes1stream}--\ref{theta1stream}) become 
\begin{eqnarray}
\nabla^4\left(\psi^{(1)}_0+\frac{1}{2}\theta^{(1)}_1\right)&=&0\label{Ereq1}\\
\partial_t\theta^{(1)}_0+\frac{1}{2}\nabla^2\psi^{(1)}_0&=&\frac{\mu}{\gamma}\nabla^2\theta^{(1)}_1.\label{Ereq2}
\end{eqnarray}
Note that since $\theta_0^{(1)}$ is harmonic, the Laplacian of Eqn.~(\ref{Ereq2}) together with Eqn.~(\ref{Ereq1}) imply that both $\psi^{(1)}_0$ and $\theta^{(1)}_1$ are biharmonic.  We immediately conclude that $\psi_0^{(1)}$ is the same as Taylor's Stokes flow solution for an isotropic liquid~\cite{taylor1951}, and then integrate Eqn.~(\ref{Ereq2}) using the anchoring condition~(\ref{bc1anch}) to find
\begin{eqnarray}
\psi_0^{(1)}&=&\varepsilon_b(1+y)\mathrm{e}^{-y}\sin(x-t)\label{fields}\\
\theta_1^{(1)}&=&-\frac{\varepsilon_b}{2}\frac{\gamma}{\mu}\frac{[1+(1+w)y]}{(1+w)^2}\mathrm{e}^{-y}\sin(x-t).\nonumber
\end{eqnarray}
Note that $\theta^{(1)}_1$ vanishes when $w\gg1$, i.e. $\theta^{(1)}$ is harmonic to first order in Er when the anchoring is strong, in accord with our large-$w$ solution~(\ref{theta1sol}). The  expressions for the fields~(\ref{fields}) yield the power dissipated to first order in Er:
\begin{eqnarray}
\frac{\mathcal{P}}{\mu\omega^2}&\approx&\int\left[2 v_{0ij}^{(1)}v_{0ij}^{(1)}+\frac{\mu}{\gamma}(\nabla^2\theta_1^{(1)})^2\right]\mathrm{d}x\mathrm{d}y\nonumber\\
\frac{P_\mathrm{T}}{\mu q\omega^2b^2}&=&1+\frac{\gamma}{4\mu}\frac{1}{(1+w)^2}+\mathcal{O}(\mathrm{Er}^2).\label{PtranslowEr}
\end{eqnarray}
In the limit of strong anchoring, $w\gg 1$, the low-Er power again goes to the isotropic limit
$\mu q\omega^2b^2$. When $\mathrm{Er}\ll1$ and the anchoring is weak, $w\ll 1$, we have $P_\mathrm{T}\approx(\mu+\gamma/4)q\omega^2b^2$. The dependence of the power dissipated on Ericksen number for various anchoring strengths and $\gamma=\mu$ is shown in Fig.~\ref{Pv}.

To find the swimming speed and flux to leading at low Ericksen number, we must expand the second-order equations~(\ref{hextf2a}--\ref{hextf2b}) in powers of Er. To zeroth order in Er we find that the equations only demand that $\langle\partial_y^2\theta^{(2)}_0\rangle=0$. Since $\langle\theta^{(2)}_0\rangle$ cannot diverge when $y\rightarrow\infty$,  $\langle\theta^{(2)}_0\rangle$ must be constant. The constant is determined by the zeroth-order terms in the anchoring condition~(\ref{anchBC2order}), which yields $\langle\theta^{(2)}_0\rangle=0$. 

To first order in Er,  Eqns.~(\ref{hextf2a}--\ref{hextf2b}) are
\begin{eqnarray}
\langle \partial^2_yv_{0x}^{(2)}\rangle
+\frac{1}{2}\langle \partial^3_y\theta^{(2)}_1\rangle&=&\langle\partial_x\theta^{(1)}_0\nabla^2\theta^{(1)}_1\rangle\label{hextf2aEr1}\\
\frac{\mu}{\gamma}\langle\partial_y^2\theta^{(2)}_1\rangle-\frac{1}{2}\langle\partial_y v_{0x}^{(2)}\rangle&=&\langle \mathbf{v}^{(1)}_0\cdot\nabla\theta^{(1)}_0\rangle,
\label{hextf2bEr1}
\end{eqnarray}
These equations along with the no-slip boundary condition lead to 
\begin{eqnarray}
\langle v_{0x}^{(2)}\rangle&&=1/2
-\frac{\gamma}{\gamma+4\mu}\frac{w}{2(1+w)^2}\times\nonumber\\
&&\left[(3+2w)(1-\mathrm{e}^{-2y})
-2(1+w)y\mathrm{e}^{-2y}\right].\label{VTlowEr}
\end{eqnarray}
The swimming speed for a transverse wave at low Er is therefore
\begin{equation}
U_\mathrm{T}=\frac{\omega q b^2}{2}\left[\frac{1}{2}-\frac{\gamma}{4\mu+\gamma}\frac{ w(2w+3)}{2(1+w)^2}\right]+\mathcal{O}(\mathrm{Er}^2).
\end{equation}
These expressions capture the low-Er asymptotic behavior depicted in Fig.~\ref{UvEr}. Likewise, the dependence of the flux on the anchoring strength at low Er (see Fig.~\ref{Qtvg}) follows from the flow field~(\ref{VTlowEr}):
\begin{equation}
Q_\mathrm{T}=\frac{1}{4}\frac{\gamma\omega b^2}{\gamma+4\mu}\frac{w(3w+4)}{(1+w)^2}+\mathcal{O}(\mathrm{Er}^2).
\end{equation}

\subsubsection{Longitudinal wave}
The analysis of the low-Er limit of the longitudinal wave case is similar to the transverse wave case. To linear order in $\varepsilon_a$, the governing equations are the same as Eqs.~(\ref{theta10a}--\ref{theta10b}) and~(\ref{Ereq1}--\ref{Ereq2}); the only change is the boundary conditions~(\ref{bc1long}--\ref{bc2long}). Solving these equations with these boundary conditions yields 
\begin{eqnarray}
\theta^{(1)}_0&=&0\\
\theta^{(1)}_1&=&-\varepsilon_a \frac{\gamma}{2\mu}\frac{1+(1+w)y}{1+w}\mathrm{e}^{-y}\cos(x-t)\\
\psi^{(1)}_0&=&-\varepsilon_a  y\mathrm{e}^{-y}\cos(x-t).
\end{eqnarray}
Once again the leading order stream function is the same as isotropic Stokes flow. 
The deformation of the swimmer does not disturb the angle field in the region adjacent to the swimmer; the disturbance to the angle field is due to the flow, and therefore $\theta^{(1)}\propto\mathrm{Er}$ for small Er. We expect $\theta^{(1)}$  or $\mathbf{v}^{(1)}$ to depend only weakly on the anchoring strength, and this expectation is reflected in the weak dependence of the power on $w$ (Fig.~\ref{PvL}). At small Er, the power dissipated by the longitudinal-wave swimmer is independent of anchoring strength, and equal to the power dissipated by the tranverse-wave swimmer (with $b=a$) with weak anchoring:
\begin{equation}
\frac{P_\mathrm{L}}{\mu q\omega^2 a^2}=1+\frac{\gamma}{4\mu}+\mathcal{O}(\mathrm{Er}^2).
\end{equation}

At second order in $\varepsilon_a$, the governing equations~(\ref{hextf2a}--\ref{hextf2b}) along with the anchoring boundary condition again imply that $\langle\theta^{(2)}_0\rangle=0$. Furthermore, since $\theta_0^{(1)}=0$, Eqns.~(\ref{hextf2a}--\ref{hextf2b})  reduce to $\langle\partial_y^2v_{0x}^{(2)}\rangle=0$. Since the first order flow to leading order in Er is Stokes flow, we conclude that to second order in $\varepsilon_a$, the swimming speed and flux are the same as in the isotropic case: 
\begin{equation}
U_\mathrm{L}=-\frac{\omega q a^2}{2}\left[1+\mathcal{O}(\mathrm{Er}^2)\right],
\end{equation}
and $Q_\mathrm{L}/(\omega a^2)=\mathcal{O}(\mathrm{Er}^2)$.


\subsection{Large Ericksen number}
When the Ericksen number is large,  the form of the decay rate $k$ [Eqn.~(\ref{keqn})] implies a boundary layer near the swimmer of thickness 
\begin{equation}
\delta\propto\sqrt{\frac{\gamma+4\mu}{\gamma}}\frac{1}{\sqrt{\mathrm{Er}}}
\end{equation}
in both the angle field and flow field, as long as $\gamma\neq0$ 
and  $c_2\neq0$ [recall that $c_0$, $c_1$, and $c_2$ are the coefficients in the solutions~(\ref{psieqn}--\ref{thetaeqn}) of the linearized equations]. The strength of the anchoring $w=W/(Kq)$ does not affect the boundary layer thickness. Inside the boundary layer, adjacent to the swimmer, elastic forces and torques balance with viscous forces and torques. Outside the boundary layer, the elastic effects can be disregarded, and the local rate of rotation of the angle field is equal to the  local rate of rotation of the fluid. The boundary layer has a small effect on the power dissipated, swimming speed, and flux. To see why, 
 we expand the exact solutions  powers of $1/\sqrt{\mathrm{Er}}$ to find for a transverse wave that 
\begin{eqnarray}
c_0&=&-\mathrm{i}+\frac{1}{2\mathrm{Er}}+\mathcal{O}(1/\mathrm{Er}^{3/2})\\
c_1&=&-\mathrm{i}+\frac{1}{2\mathrm{Er}}+\mathcal{O}(1/\mathrm{Er}^{3/2})\\
c_2&=&\frac{1}{k}-\mathrm{i}\frac{w}{\mathrm{Er}}\frac{4\mu+\gamma}{4\gamma}+\mathcal{O}(1/\mathrm{Er}^{3/2}).
\end{eqnarray}
Recall that $k\propto\sqrt{\mathrm{Er}}$ for $\mathrm{Er}\gg1$. Since $w$ enters in a term that is subleading in Er, we see that the effects of anchoring vanish at larger Ericksen number. Note that to zeroth order in Ericksen number, we have $c_0=c_1=-\mathrm{i}$, and $c_2=0$. In other words, to leading order in Ericksen number, the solution to first-order in dimensionless amplitude $\varepsilon_b$ is precisely the strong-anchoring solution~(\ref{theta1sol}--\ref{p1sol}). The strong-anchoring solution is the `outer' solution, valid outside the boundary layer, $y\gtrsim\delta$. Within the boundary layer, the angle field rapidly changes from the strong-anchoring condition to whatever value is necessary to satisfy the anchoring condition (\ref{anchBC}) for finite $w$. However, since $c_2$ gets smaller and smaller as the boundary layer gets thinner and thinner, the effects of the boundary layer on the problem is small. Thus, for large Ericksen number, the power dissipated, swimming speed, and flux are given by the strong-anchoring limit values: $P_\mathrm{T}\sim\mu q\omega^2b^2$, $U_\mathrm{T}\sim U_\mathrm{Ta}$, and $Q_\mathrm{T}\sim Q_\mathrm{Ta}$. 
The same reasoning also applies to the longitudinal case, where for large Er we find
$P_\mathrm{L}\sim\mu q\omega^2a^2$,
$U_\mathrm{L}\sim-\omega q a^2/2$, and
$Q_\mathrm{L}\sim{\gamma\omega a^2}/({16\mu+4\gamma})$.

\section{Summary} We have calculated the flow field in a two-dimensional hexatic liquid crystal generated by an infinite one-dimensional swimmer with internally generated transverse or longitudinal traveling waves. Working to second order in the amplitude, we found the power dissipated, the swimming speed, and the fluid pumped by the swimmer. For a transverse wave, the swimming speed and power dissipated depends strongly on the rotational viscosity $\gamma$ for all Ericksen numbers, and on the anchoring conditions for low Ericksen number, which is expected to be the relevant regime for real swimming microorganisms. For a longitudinal wave, the swimming speed and power generated is virtually identical to that for swimming in an isotropic fluid, despite the fact that the flow differs from Stokes flow at large Ericksen number. For both kinds of waves, there is a nonzero flux of fluid pumped by the swimmer, in contrast to the case of a swimmer in a isotropic Newtonian fluid or a viscoelastic fluid described by the  Oldroyd-B model.

We found that the swimmer causes a uniform disturbance in the angle field infinitely far from the swimmer.  This result likely arises as an artifact of our sinusoidal-steady state assumption; in a future publication we will examine the `start-up' problem in which the swimmer starts from rest and accelerates to its steady speed~\cite{PakLauga2010}.  It will also be interesting to generalize our calculations to other geometries, such as circular or spherical squirmers~\cite{Lighthill1952} and helical flagella. Likewise, it will be important to generalize these calculations for swimmers in two- and three-dimensional nematic liquid crystals. \\






\begin{acknowledgments}
This work was supported in part by National Science Foundation Grant No. CBET-0854108 (TRP) and CBET-1437195 (TRP).
Some of this work was carried out at the Aspen Center for Physics, which is supported by National Science Foundation Grant No. 1066293. We are grateful to John Toner for insightful comments and advice at the early stages of this work, and to Marcelo Dias for discussion.
\end{acknowledgments}


\bibliography{../../../../newrefs}

\end{document}